# Tethered capsule *en face* optical coherence tomography for imaging Barrett's esophagus in unsedated patients

Kaicheng Liang, Osman O. Ahsen, Annalee Murphy, Jason Zhang, Tan H. Nguyen, Benjamin M. Potsaid, Marisa Figueiredo, Qin Huang, Hiroshi Mashimo, James G. Fujimoto

**Kaicheng Liang, Osman O. Ahsen, Jason Zhang, Tan H. Nguyen, Benjamin M. Potsaid, James G. Fujimoto**: Department of Electrical Engineering and Computer Science, and Research Laboratory of Electronics, Massachusetts Institute of Technology, Cambridge MA, USA

**Annalee Murphy, Marisa Figueiredo, Qin Huang, Hiroshi Mashimo**: VA Boston Healthcare System, Boston MA, USA

**Qin Huang, Hiroshi Mashimo**: Harvard Medical School, Boston MA, USA

**Author Contributions**

KCL, JGF, HM designed the study; KCL, OOA, BMP developed the OCT imaging technology; KCL, OOA, AM, JZ, THN, MF, HM collected the data; KCL, OOA, JGF, HM analyzed the data; QH made the histological diagnoses; JGF and HM obtained funding for the study; KCL, JGF and HM wrote the manuscript; All authors read and contributed to the manuscript; JGF and HM were principal investigators for this study.

**Financial support**

National Institutes of Health grants R01-CA075289-21 (JGF and HM) and R44CA235904-02 (JGF), graduate fellowship from Agency for Science, Technology and Research, Singapore (KCL).

**Corresponding author:** James G. Fujimoto PhD, Department of Electrical Engineering and Computer Science, and Research Laboratory of Electronics,

Massachusetts Institute of Technology, Cambridge MA, USA jgfuji@mit.edu

Tel: +1-617-253-8528

Main text word count: 3336

The Submitting Author accepts and understands that any supply made under these terms is made by BMJ to the Submitting Author unless you are acting as an employee on behalf of your employer or a postgraduate student of an affiliated institution which is paying any applicable article publishing charge ("APC") for Open Access articles. Where the Submitting Author wishes to make the Work available on an Open Access basis (and intends to pay the relevant APC), the terms of reuse of such Open Access shall be governed by a Creative Commons licence – details of these licences and which Creative Commons licence will apply to this Work are set out in our licence referred to above.

**Abstract** (up to 300 words)

Objective

Detection of Barrett's esophagus (BE) at points of care outside the endoscopy suite may improve screening access and reduce esophageal adenocarcinoma mortality. Tethered capsule optical coherence tomography (OCT) can volumetrically image esophageal mucosa and detect BE in unsedated patients. We investigated ultrahigh-speed tethered capsule, swept-source OCT (SS-OCT) in unsedated patients, improved device design, developed procedural techniques, measured how capsule contact and longitudinal pullback non-uniformity affect coverage, and assessed patient toleration.

Design

OCT was performed in 16 patients prior to endoscopic surveillance/treatment. Unsedated patients swallowed the capsule with small sips of water and the esophagus imaged by retracting the tether. Ultrahigh-speed SS-OCT at 1,000,000 A-scans/second imaged ~40cm$^2$ esophageal areas in 10 seconds with 30μm transverse and 8μm axial resolution. Capsule contact and longitudinal image coverage were analyzed. Patients were interviewed to assess toleration.

Results

Nine patients had non-dysplastic BE, 3 had ablative treatment-naïve neoplasia, and 4 had prior ablation for history of dysplasia. Dry swallows facilitated capsule transit through the lower esophageal sphincter (LES), and waiting 10 seconds before retrieval reduced swallow-induced LES relaxation. Slow nasal inhalation facilitated capsule retrieval and minimized the gag reflex. The capsule procedure was well tolerated. Ultrahigh-speed SS-OCT enabled the generation of cross-sectional and sub-surface *en face* images. BE could be rapidly identified in cross-sectional and *en face* images, while assessment of the gastro-esophageal junction (GEJ) required subsurface *en face* views. Capsule-esophageal contact was worse in long-segment BE and sliding hiatal hernia, and longitudinal image coverage was worse in short-segment BE. OCT candidate features of dysplasia are reported.

Conclusions

Ultrahigh-speed tethered capsule SS-OCT enabled *en face* and cross-sectional imaging of mucosal features over wide areas. Device and procedure optimization led to improved imaging performance. Areas of BE could be readily identified, but limited capsule contact and longitudinal image coverage can yield sampling errors.

**Summary Box** (separate from abstract)

*What is already known about this subject?*

Modalities for Barrett's esophagus screening and risk stratification outside the endoscopy suite can improve access to care and potentially reduce mortality. Previous optical coherence tomography (OCT) imaging studies with balloons and tethered capsules have shown diagnostic potential for BE and dysplasia, but acquire sparse cross-sectional images with limited resolution and coverage. Procedural information on tethered capsule OCT has been limited.

*What are the new findings?*

Ultrahigh-speed SS-OCT tethered capsules can generate *en face* and cross-sectional images, mapping wide areas of the esophagus and gastroesophageal junction, enabling quantitative assessment of surface and subsurface features in patients without requiring endoscopy or sedation. Device design and procedural techniques are described for optimizing examination performance in unsedated patients.

*How might it impact on clinical practice in the foreseeable future?*

Next-generation, ultrahigh-speed SS-OCT tethered capsule imaging may improve screening for Barrett's esophagus and risk stratification at points of care outside the endoscopy suite.

**Introduction**

Barrett's esophagus (BE) and dysplasia are precursors to esophageal adenocarcinoma (EAC). However, 90% of EAC patients never receive an endoscopic BE diagnosis[1] and 40% do not report symptoms of chronic gastroesophageal reflux disease[2]. Therefore new BE screening methods that can be used at points of care outside the endoscopy suite are needed to improve early detection of BE before progression to EAC. The Cytosponge has been validated for BE detection in large trials with 80% sensitivity and 92% specificity[3], and evaluated for risk stratification[4]. Unsedated transnasal endoscopy has been compared to endoscopy, showing comparable detection of BE, although transnasal intubation requires expertise and tolerability results have been mixed[5 , 6]. Breath testing has demonstrated ~80% sensitivity/specificity to BE[7]. DNA methylation markers and a swallowable device for tissue sampling achieved >90% sensitivity/specificity to BE[8].

Endoscopic optical coherence tomography (OCT) can image subsurface tissue architectural morphology and early studies suggested its potential for BE and dysplasia detection[9]. Balloon-based OCT has been commercialized as volumetric laser endomicroscopy (VLE, NinePoint Medical)[10]. Tethered capsules can be swallowed without sedation; early string-tethered video capsules showed sensitivity and specificity of 93.5% and 78.7% compared with histological BE diagnosis[14], and were later adapted for OCT imaging[15]. OCT tethered capsules can image long lengths of the esophagus by slow retraction of the tether after swallowing. OCT capsule studies demonstrated cross-sectional images, reported excellent patient toleration[16] and showed correlations with endoscopic Prague measurements[17]. However, there are limited investigations of capsule contact with the esophagus and non-uniform longitudinal capsule motion, which reduce image coverage and cause sampling errors. Procedural details for tethered OCT capsule imaging in unsedated patients have also been limited.

Our group previously reported ultrahigh-speed OCT using micromotor probes introduced into the endoscope working channel that generate *en face* and cross-sectional views enabling 3-dimensional assessment of tissue architectural morphology[18]. Using micromotor imaging probes, we reported *en face* and cross-sectional OCT features associated with BE and dysplasia[19]. We also evaluated the feasibility of ultrahigh-speed, micromotor tethered capsule OCT for volumetric and *en face* imaging >20cm lengths of the esophagus in sedated patients during endoscopy, demonstrating large field-of-view, detecting BE, and suggesting features associated with dysplasia[20]. Here we report ultrahigh-speed, tethered capsule OCT for mapping the esophagus and imaging BE in unsedated patients. We describe device improvements, procedural techniques for optimal imaging, measure capsule contact, and longitudinal image coverage for BE detection, and assess patient toleration. This information is relevant for future study design and wider-spread screening applications.

**Methods**

*Imaging system and tethered capsule*

Tethered capsule OCT imaging was performed using a prototype ultrahigh-speed swept-source OCT (SS-OCT) instrument operating at 1,000,000 A-scans/second, 20x faster than commercial endoscopic OCT (VLE NinePoint). The tethered capsules (Figure 1A-B) used micromotors for circumferential imaging at 300 cross-sectional images per second[20] with 30µm transverse and 8µm axial resolution. Several improvements were made over previously reported capsule designs[16]. Our new capsule was 12mm diameter with proximal and distal ends made with lubricious, medical-grade ultrahigh-molecular-weight polyethylene and a small polycarbonate transparent window for the OCT beam. This new design reduced friction, allowing the capsule to be swallowed and retrieved smoothly. The proximal end had a 30° taper, improving pullback smoothness and ease of retrieval through the lower esophageal sphincter (LES) and upper esophageal sphincter (UES). The tether was 2.2mm diameter, providing flexibility for patient comfort while retaining some rigidity for operator control, and was marked at 5cm intervals to assess capsule distance from the incisors.

*Patient recruitment and imaging procedure*

The study was approved by IRBs at the Veterans Affairs Boston Healthcare System, Harvard Medical School, and Massachusetts Institute of Technology. Written informed consent was obtained from patients undergoing BE surveillance or endoscopic treatment for prior diagnosed neoplasia. Neoplasia was defined to include low-grade dysplasia (LGD), high-grade dysplasia (HGD), and intramucosal carcinoma (IMC). Patients were scheduled for same-day sedated endoscopy and underwent standard endoscopy preparation.

Prior to sedation and endoscopy, patients swallowed the tethered capsule with small sips of water while sitting upright. The procedure was supervised by an endoscopist who controlled the tether. Wet and dry swallows facilitated capsule transit into the stomach, and the capsule was positioned at the gastric cardia, as monitored using real-time OCT imaging. The endoscopist then imaged a ~10cm long segment of the esophagus by pulling back the tether at ~1cm/sec for ~10sec, causing the capsule to move longitudinally through the GEJ (Figure 1C). The pullback length was estimated by the tether markings. Dense volumetric datasets with ~10 million A-scans and ~5 gigavoxels, covering a ~40cm$^2$ area of the esophagus were acquired (Figure 1D).

Patients were asked to take dry swallows to improve capsule contact with the esophageal wall, and the endoscopist waited ~10 sec prior to the pullback in order to avoid swallow-induced LES relaxation (deglutitive inhibition)[21]. Imaging was performed during pullback and not during the capsule's natural migration after swallows because peristalsis, generally progressing at several cm/sec, is too rapid for dense volumetric imaging. Capsule contact with the esophagus was assessed by observing the cross-sectional image series and *en face* images. If substantial out of contact regions were noted, additional dry/wet swallows were performed to improve contact before repeating pullback image acquisitions. If excessive longitudinal capsule motion non-uniformity occurred, as indicated by longitudinal stretching or compression of features in the *en face* image, additional pullback acquisitions were repeated after a brief 10-15 sec wait post-swallow to avoid the rebound LES contractions. After imaging, the capsule was retrieved by retracting the tether. Swallowing motions were not effective for capsule retrieval because the relaxation of the UES is transient and this places the posterior tongue closer to the tether, causing gagging. Instead, patients were asked to perform slow nasal inhalation, minimizing gag reflex from the posterior tongue.

Patients then underwent sedated endoscopy ~2-4 hours later, following the American Society for Gastrointestinal Endoscopy guidelines requiring 2-hour fasting after ingestion of clear liquids before intravenous sedation[22]. During endoscopy, measurements were obtained of BE length and diaphragmatic hiatus. Seattle protocol 4-quadrant biopsy was performed in non-dysplastic BE (NDBE) surveillance patients, and endoscopic treatment performed in patients with prior diagnosed dysplasia, per standard of care. In patients referred for treatment with known prior treatment-naïve dysplasia, biopsies/resections were not systematically obtained unless clinically indicated.

After tethered capsule OCT and endoscopy, a research nurse performed a brief patient interview to assess toleration for unsedated tethered capsule OCT and sedation endoscopy. The post-procedure patient interview questions were: "How anxious did you feel before the procedure? 1 not, 5 very", "How much discomfort did you have during the procedure? 1 none, 5 a lot", and "Would you recommend the procedure to others? 1 definitely yes, 5 definitely no". The questions were identical to those reported in previous OCT capsule imaging studies in order to facilitate comparison[16].

*Data analysis*

Datasets were assessed for image quality and one optimal dataset from each patient was selected for analysis. The region of analysis on the *en face* OCT images was defined to be the portion of the esophageal wall that closely encircled the capsule, with proximal margin at the squamocolumnar junction (SCJ) maximal extent and longitudinal extent as observed by endoscopy (Prague M length). The endoscopically observed BE length was used to help delineate the extent of BE in the *en face* OCT images because the GEJ is a morphological transition zone that could not be reliably delineated by OCT features alone. Since this analysis was retrospective, endoscopic measurements aided the OCT assessment. However, this suggests that quantitative measurements of BE length (Prague M length) using OCT alone will be challenging because of uncertainty in GEJ location. Analogously, endoscopic identification of the GEJ relies on discerning the top of the gastric folds, a gross anatomical landmark that is similarly prone to uncertainty.

The longitudinal capsule motion was non-uniform compared with the tether pullback, therefore regions, where the capsule moved more slowly/rapidly, could be detected as stretching/compression of features in the longitudinal direction of the *en face* OCT images. This pullback non-uniformity caused distortion or gaps in the OCT data, limiting longitudinal image coverage, producing sampling errors. This also made quantitative measurements of BE length (Prague M length) challenging.

Contact of the capsule with the esophagus and longitudinal motion non-uniformity were analyzed as metrics for image coverage and sampling error. Capsule-tissue contact was measured as the percent of the capsule circumference where the esophagus was within ~100μm from the capsule surface. The mean of the capsule-tissue contact was calculated from the region in the *en face* OCT from the GEJ to the maximal extent of visible BE at the SCJ. Squamous mucosa regions proximal to the SCJ maximal extent were excluded from the analysis, because these areas usually showed good contact, but were less relevant to BE assessment.

Longitudinal image coverage was assessed by identifying longitudinally stretched regions in the *en face* OCT where the capsule stopped or moved too slowly relative to the esophagus during pullback. The percentage of the image with longitudinal stretching was calculated. However, areas where the capsule moved too fast relative to the esophagus appeared as longitudinally compressed regions in *en face* OCT. These regions comprise a negligible portion of the total longitudinal image but represent the sampling error. The image data is displayed as a function of time. The regions with overly slow motion were assumed to be approximately equal to the regions with overly fast motion because the tether pullback speed is constant and therefore the overall pullback length and time are known. Therefore we used the longitudinally stretched areas in the *en face* OCT as a marker for estimation of longitudinal image coverage and sampling error.

Tissue contact and longitudinal image coverage were stratified by short/long-segment BE (<=3cm and >3cm), and absence/presence of sliding hiatal hernia (the distance of diaphragmatic hiatus from gastric folds <=2cm and >2cm) subgroups.

The patient toleration (anxiety and discomfort) and recommendation scores were analyzed. The toleration score distributions were stratified by pathology and treatment subgroups to show possible associations with prior endoscopy and/or treatment experience. Statistical analysis was not performed due to the small enrollment size and the observational nature of the study.

**Results**

*Patient demographics and clinical characteristics*

Sixteen patients were enrolled. Table 1 summarizes patient demographics including baseline pathology (neoplasia status) at the time of imaging, treatment history, and BE length. A mean of 48±27 milliliters of water was consumed during the tethered capsule procedure. In all but one patient, tethered capsule imaging was followed by an esophagogastroduodenoscopy (EGD). This patient was scheduled for an unrelated colonoscopy, but was in ongoing surveillance for long-segment BE, with a recent EGD in the previous year. Therefore, this patient was not asked to provide toleration scores for endoscopy.

*Tissue contact with capsule and longitudinal coverage*

A mean of 5.5±1.3 pullback datasets were obtained per patient, from which an optimal dataset was selected. In 16 patients, there was a 75±27% mean percent of tissue-capsule contact averaged over the *en face* BE region of analysis. Longitudinal image coverage was 59±34% of the BE segment. Tissue contact was associated with endoscopic BE length and sliding hiatal hernia (Figure 2). The mean tissue contact was 89±11% for short-segment BE with/without prior ablative treatment (n=8) patients, and 61±31% for long-segment BE (n=8) (p=0.03). The mean tissue contact was 84±15% in patients without sliding hiatal hernia (n=11), and 55±37% with hernia (n=5) (p=0.04). Folds in the esophagus sometimes occurred, suggesting that the percent of esophagus which was not imaged exceeds the percent of capsule circumference which was out of contact. Also, BE is not fully circumferential in patients with Prague M>C. Therefore, the percentage of BE imaged may be greater or less than that inferred from capsule contact.

*Toleration scores*

The mean procedure time for tethered capsule imaging was 9.7±3.0 minutes. For capsule imaging (n=16), the mean pre-procedure anxiety was 1.9±1.0, procedural discomfort was 2.5±1.1, and recommendation score was 1.3±0.7. For endoscopy (n=15), the mean pre-procedure anxiety was 1.3±0.7, procedural discomfort was 1.6±0.9, and recommendation score was 1.1±0.3. Toleration did not appear to be associated with the number of imaging pullbacks performed. The toleration score distributions, grouped by patient history of pathology/treatment are presented in Figure 3.

Toleration scores stratified by pathology and treatment history were generally consistent between subgroups, suggesting that toleration may be independent of prior endoscopy and treatment experience. Endoscopy was better tolerated than the tethered capsule, likely due to sedation. Our toleration scores show marginally lower pre-procedure anxiety (1.9±1.1 vs 2.1±0.8) and higher procedural discomfort (2.5±1.0 vs 1.9±0.9) than previous reports[16], possibly due to our larger capsule diameter (12 mm vs 11 mm) chosen for better tissue contact. These results suggest that tethered capsule OCT is well-tolerated, although both our study and the previous study were single-center and our study enrollment was all male veterans, so results are not generalizable. There were no adverse events during the entire study.

*OCT features of BE*

Figure 4 shows a wide-field *en face* OCT over an entire BE segment from GEJ to SCJ. This data is from the patient who did not receive same-day EGD but had known long-segment NDBE from two EGDs 9 months prior and 4 years prior (4-quadrant biopsies found no dysplasia). OCT showed almost complete contact of the esophageal mucosa to the capsule, while some loss of longitudinal coverage from longitudinal capsule motion non-uniformity was observed.

Numerous dilated glands were observed over the entire BE segment, and an exceptionally high number of gland clusters was observed on the gastric side of the GEJ. This example showed exceptionally well-defined boundaries of the gastroesophageal junction (GEJ) and squamocolumnar junction (SCJ). This example also illustrates that important OCT markers may frequently occur at the gastric cardia around the GEJ, even in low-risk NDBE patients[23]. Features at the GEJ can confound attempts to identify neoplasia, particularly if the GEJ boundary is poorly defined and wide-field *en face* OCT views are unavailable. We report candidate OCT features for dysplasia in the Supplementary Information.

**Discussion**

We report a pilot study using ultrahigh-speed, tethered capsule SS-OCT imaging for wide-field, *en face* and cross-sectional visualization. Previous studies with commercial OCT instruments used slower, 50,000 A-scans per second, imaging speeds and lower, 40µm transverse resolution, generating sparsely-spaced, cross-sectional image volumes that did not have high resolution *en face* views. Our OCT technology at 1,000,000 A-scans/sec and 30µm resolution can generate densely sampled volumes enabling *en face* imaging over wide fields of view. Previous OCT capsule studies investigating screening recruited largely from a primary care population, whose patients had little or no BE, and those study designs did not ensure same-day endoscopy for direct comparison. In this study, we recruited from a heterogeneous BE surveillance population in which patients had various BE lengths and history of dysplasia. Improved capsule design using lubricious materials and tapered form factor, improved procedural workflow, and increased capsule imaging performance. This study assessed procedures and performance for imaging unsedated patients prior to endoscopy, however, tethered capsule imaging can also be performed during endoscopy.

Limitations of this study include the lack of histological correlation with OCT features. 4-quadrant biopsies were performed in patients with NDBE history, but patients with prior treatment-naïve dysplasia and referred for treatment were not systematically biopsied. We performed additional analyses of OCT features that may be associated with dysplasia (Supplementary Information), with the caveat that these observations had histological corroboration on per-patient basis, but not at the precise location where the OCT features were found. Our capsule did not have the capability to mark areas of interest for biopsy or to obtain biopsy directly. However, laser marking based on cross-sectional image guidance using tethered capsules has been reported[24]. Future capsule studies could adapt the laser marking paradigm to guide biopsy based on both *en face* and cross-sectional features. Another limitation was the small patient enrollment from a single-center, therefore results on toleration, image coverage and image quality may not generalize to larger studies. However, this type of pilot study is important for optimizing device design and examination protocols; identifying candidate OCT markers of BE, dysplasia and risk stratification; and designing future large prospective studies.

Close contact of the capsule with the esophagus is critical for OCT visualization, assuring that the circumference of the esophagus is imaged and within the optical focus depth-of-field. Reduced esophageal contact can introduce optical aberrations that reduce OCT visibility of mucosal microstructure, and esophageal folds can obscure areas from view. These effects reduce image coverage and produce sampling errors. Previous studies suggested that swallow-initiated peristaltic contractions would ensure contact between capsule and esophagus. However, in our experience, patient swallows were helpful, but a substantial minority of patients had poor contact despite repeated dry/wet swallows and avoiding deglutitive inhibition of the LES. The short-segment BE patients (n=8) showed superior tissue contact compared to long-segment BE patients (n=8) (contact 89±11% vs 61±31%, p=0.03) (Figure 3). Patients without sliding hiatal hernia (n=11) showed superior contact compared to those with hernia (n=5) (contact 84±15% vs 55±37%, p=0.04). Early studies of esophageal motility in BE found associations between long BE lengths and reduction of LES tone and peristaltic amplitude[25]. The reduced LES tone associated with sliding hiatal hernia[21] may also contribute to poor contact. In the future, contact might be improved using an articulating mechanism to appose the capsule to the esophageal wall.

Longitudinal capsule motion uniformity is necessary for high-quality *en face* visualization. The capsule moved slower/faster than the tether pullback during portions of the OCT acquisitions, resulting in a localized stretched/compressed appearance in *en face* images and sampling errors. Previous studies with micromotor imaging probes[18] used 2mm/sec motorized pullback of the scanning optics within a transparent sheath to acquire volumetric OCT. Our tethered capsule was manually pulled back at ~1cm/sec while in contact with the esophagus. In our previous study using a tethered capsule during sedated endoscopy[20] capsule motion was non-uniform at slow speeds due to friction. The lubricious housing used in the current capsule reduced friction and the faster 1cm/sec pullback improved capsule longitudinal motion uniformity.

During pullback, the operator reported occasional resistance from anatomic variations, peristalsis, LES tone and/or rebound contractions, leading to non-uniform pullback. The operator did not forcefully pull through these resistances, which are meaningful anatomical signals for safe and effective use. Short-segment BE had more non-uniform capsule motion compared with long-segment BE, possibly because the contact produced by strong LES tone increased friction. Conversely, long-segment BE had looser tissue-capsule contact, which may enable smoother capsule motion. Smooth pullback and capsule-tissue contact generally showed opposite (inverse) trends, because contact produced friction. Non-uniform capsule longitudinal motion produces stretching/compression artifacts in the *en face* image, however, distortion occurs only in the pullback direction and image features can still be interpreted by experienced readers. Future attempts to perform more precise quantitative measurements or reduce sampling error in *en face* OCT images should incorporate more sophisticated motion artifact correction or motion tracking. Further increases in imaging speed will be possible and will enable faster pullbacks with reduced friction and improved longitudinal image coverage and uniformity, although pullback speeds may need to be moderated while traversing constrained sphincters.

Limitations in capsule contact produce sampling error and suggest that the current version of the capsule would not be suitable as an unsedated imaging modality for low-cost surveillance of focal pathologies such as dysplasia. However, capsules can still be used for screening because sampling errors would not appreciably compromise detection sensitivity for BE, although other methods for screening are also promising[3 , 5 , 8]. For surveillance applications, a capsule imaging device might be used during sedation endoscopy. The capsule could be attached to the endoscope (similar to a focal RFA catheter) and articulated to contact the esophageal wall, mapping longitudinal extent by advancing or retracting the endoscope. This protocol should enable comprehensive imaging coverage as well as access to the GEJ, areas of hiatal hernia, and gastric cardia.

The ultrahigh-speed tethered capsule produces better images than balloon OCT devices, should have better image coverage and could be multi-use, reducing cost.

Ultrahigh-speed tethered capsule OCT can visualize mucosal architectural morphology over wide fields-of-view. This technology may help the early screening of BE and risk stratification. Additional larger studies with histological correlation are warranted.

**Tables and Figures**

| | |
|---|---|
| Age, mean (± SD) | 68 (7) |
| Sex, male, no. (%) | 16 (100) |
| Race, white, no. (%) | 16 (100) |
| Baseline pathology and treatment status | |
|     NDBE subjects, no. (%) | 9 (56) |
|         Short-segment (<=3cm) Barrett's Esophagus (BE), no. (%) | 2 (13) |
|     LGD subjects, no. (%) | 4 (25) |
|         Ablative treatment-naïve subjects, no. (%) | 1 (6) |
|             Short-segment BE, no. (%) | 1 (6) |
|         Treated subjects, no. (%) | 3 (19) |
|             Residual short-segment BE, no. (%) | 3* (19) |
|     HGD/IMC subjects, no. (%) | 3 (19) |
|         Ablative treatment-naïve subjects, no. (%) | 2† (13) |
|             Short-segment BE, no. (%) | 1 (6) |
|         Treated subjects, no. (%) | 1 (6) |
|             Residual short-segment BE, no. (%) | 1 (6) |
| Length of BE at study endoscopy, cm | |
|     Circumferential extent, mean (± SD) | 3.6 (4.3) |
|     Maximal extent, mean (± SD) | 5.1 (4.5) |
|     Short-segment (<=3cm) subjects, no. (%) | 8 (50) |
| Long-segment (>3cm) subjects, no. (%) | 8 (50) |
| Distance from diaphragmatic hiatus (D) to gastric folds (G), mean (± SD) | 2.3 (2.5) |
| Subjects with sliding hiatal hernia (D-G >2cm), no. (%) | 5 (31) |
| Length of hiatal hernia, mean (± SD) | 5.6 (2.1) |

*One treated LGD patient had no visible BE on endoscopy and was classified as short-segment BE.

†One HGD/IMC patient had prior endoscopic mucosal resection and no ablation, thus classified as ablative treatment-naive.

Table 1. Patient demographics and clinical characteristics (n=16).

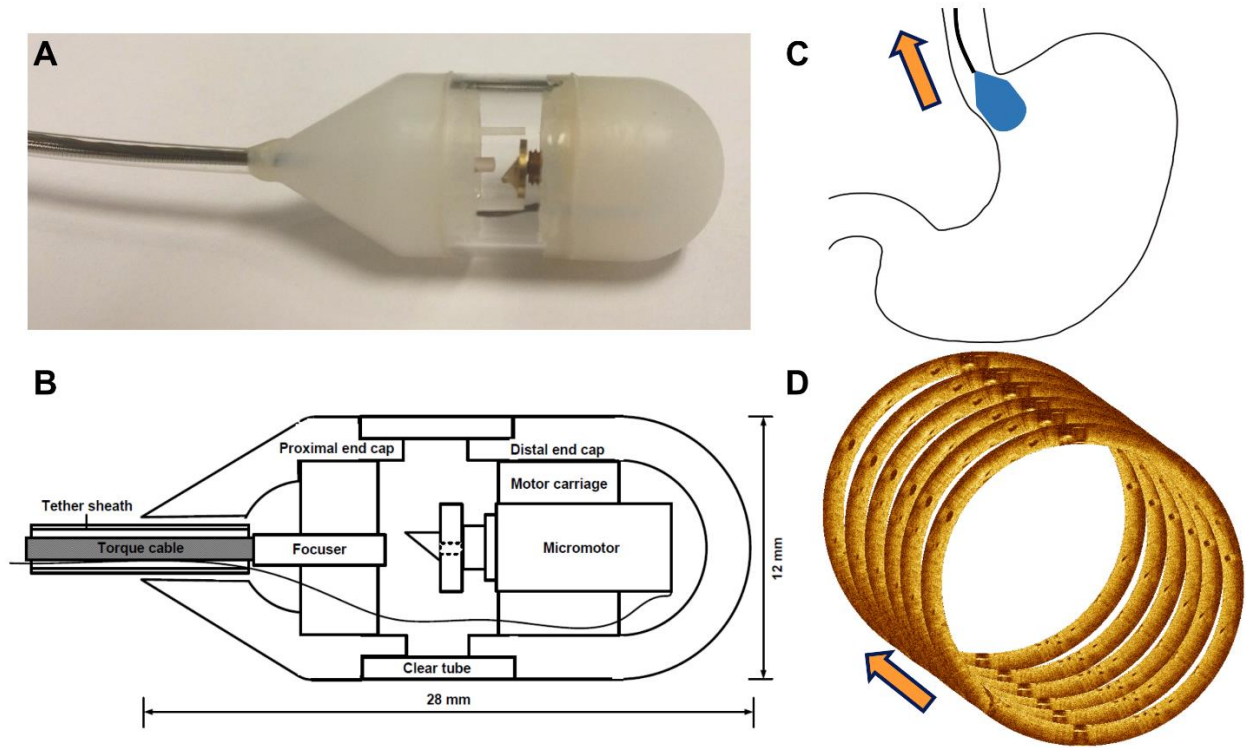

Figure 1. A: Photograph of tethered OCT capsule constructed using lubricious material with a 30° proximal taper for ease of retrieval. B: Schematic showing micromotor rotary optical scanner and other components. C: Cartoon showing capsule traveling from gastric cardia into distal esophagus during a pullback scan. D: Illustration showing multiple cross-sectional images rapidly acquired in rapid succession during capsule pullback to obtain volumetric data for subsurface *en face* and cross-sectional visualization.

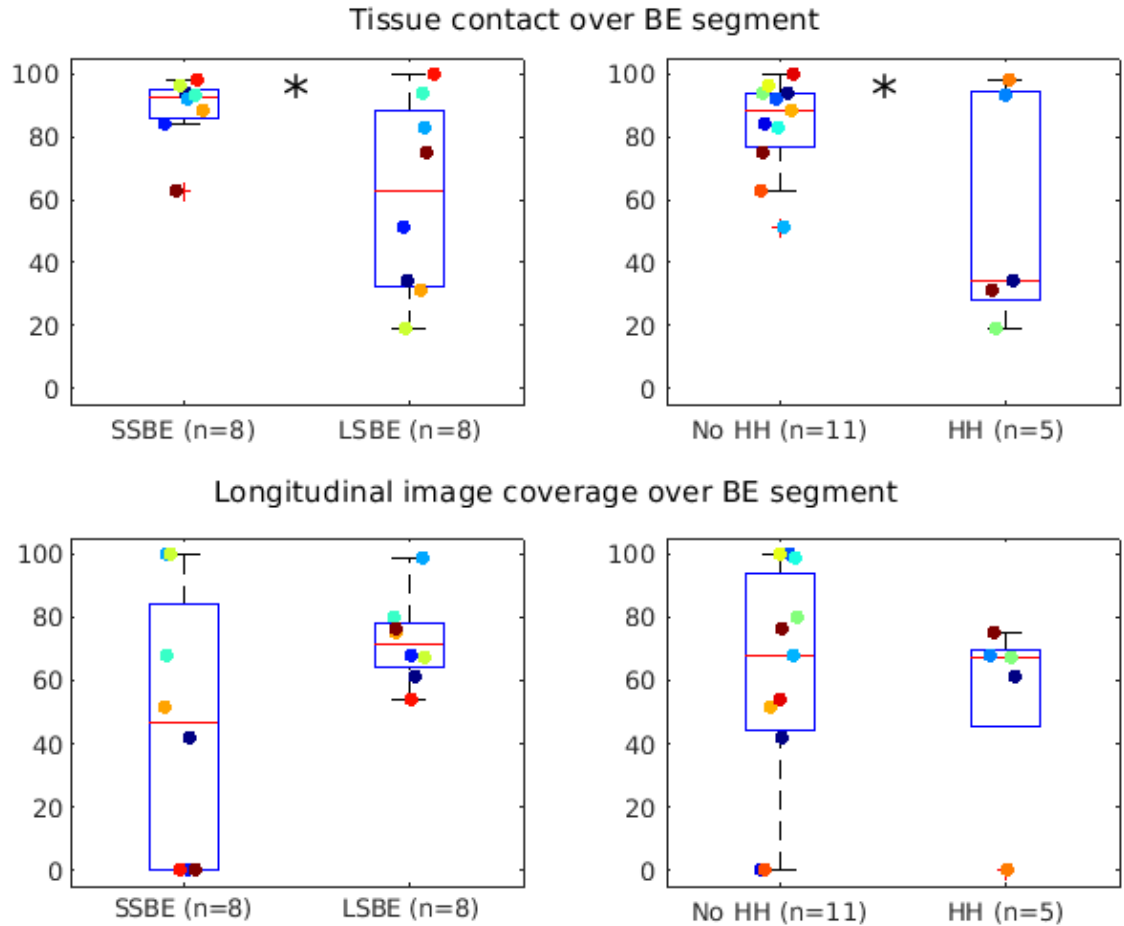

Figure 2. Boxplots of tissue contact and longitudinal capsule motion uniformity/coverage over the *en face* BE region in the tethered capsule OCT datasets. Colored circles indicate individual data points. Tissue contact was significantly different (*) between short/long segment BE (p=0.03) and absence/presence of sliding hiatal hernia (p=0.04).

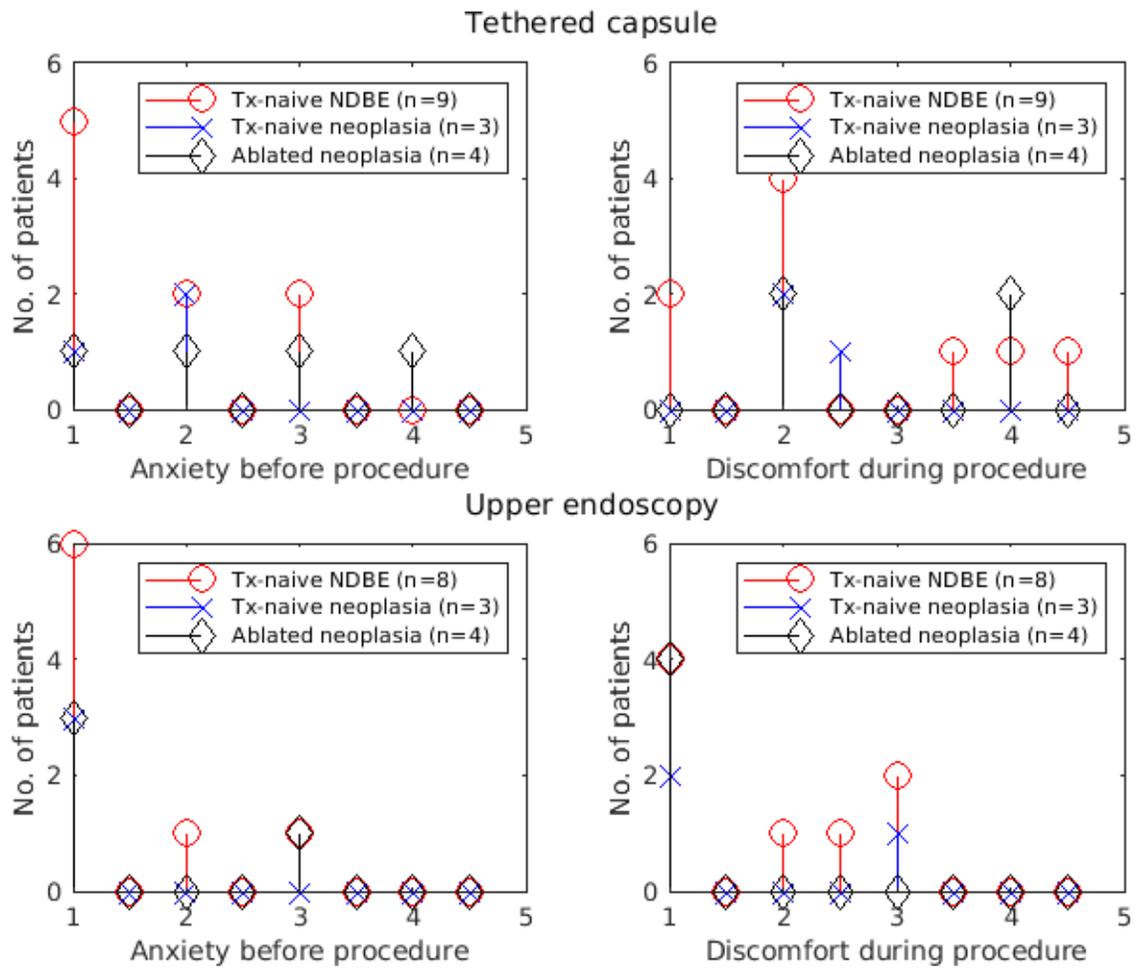

Figure 3. Pre-procedure anxiety and procedural discomfort scores for the tethered capsule and endoscopy procedures. 1- no anxiety/discomfort, 5- high anxiety/discomfort. Scores between patient subgroups of baseline pathology and treatment history were similar.

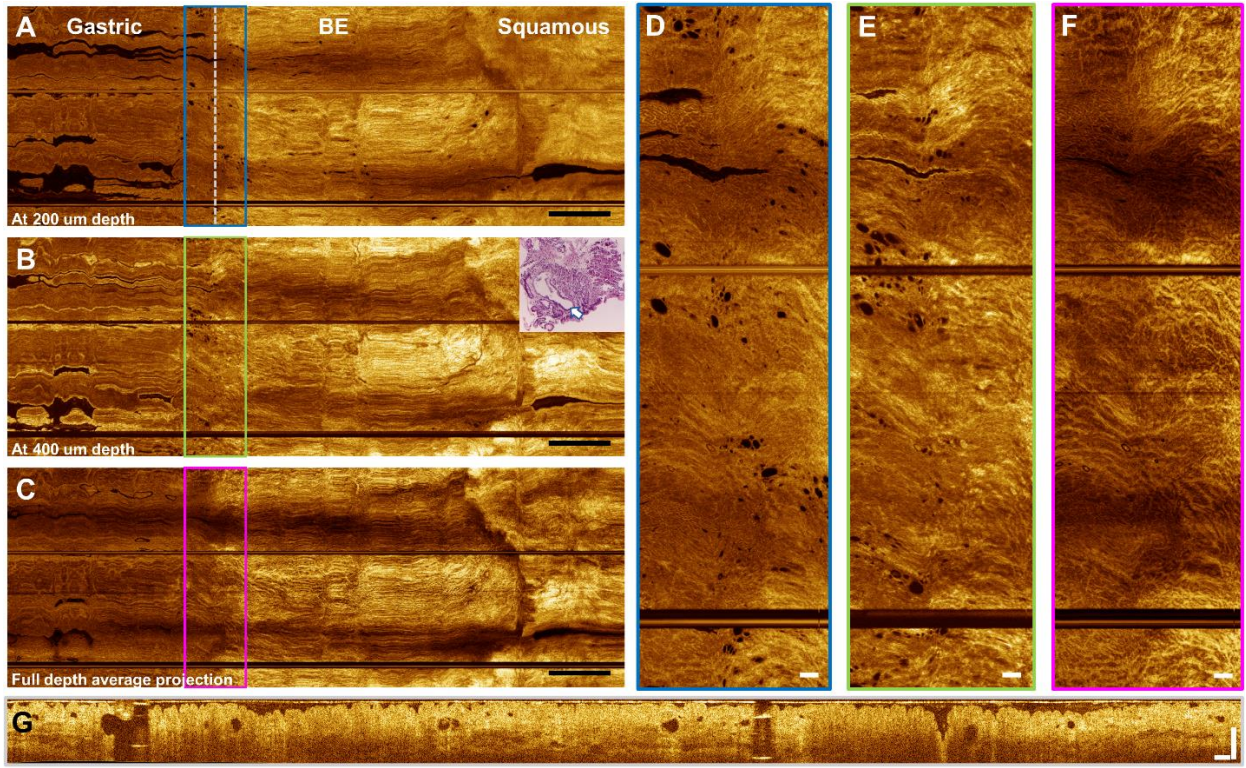

Figure 4. Tethered capsule OCT from a patient with C2M4 non-dysplastic BE. (A) en face OCT at 200 μm depth, (B) 400 μm depth, and (C) full depth projection. Scale bars 1 cm. Some longitudinal capsule motion non-uniformity can be observed in the BE segment. (D-F) Enlargements showing glands and mucosal pattern at the GEJ. Scale bars 1 mm. (G) Cross-sectional OCT from the GEJ showing atypical glands. Scale bar 500 um. Biopsy at the GEJ (inset) from a prior EGD (9 months earlier) shows a large dilated cardiac gland (arrow) with smaller peripheral glands from the superficial mucosa.

Supplementary Information

**OCT features of dysplasia**

Tethered capsule OCT enables wide-field imaging and can potentially identify focal regions of dysplasia in non-dysplastic Barrett's esophagus (NDBE). Here, we present preliminary findings that suggest candidate markers for dysplasia which could be tested in future studies.

Methods

The tethered capsule OCT datasets were assessed by an expert reader (KCL) for candidate features of dysplasia. The reader was unblinded to patient history and endoscopic findings in order to identify distinctive image features/traits associated with clinical history and biopsy histology. Two candidate features were identified and assessed: atypical gland clusters (AGCs)[1] and irregular mucosal patterns (IMPs)[2]. AGCs were defined as atypical glands, including non-elliptical shape, branching or internal debris, with a clustered density of >5 atypical glands appearing over the depth of BE in a 5mm square (25mm$^2$) *en face* area, observable in *en face* images down to ~1mm deep in BE mucosa. The *en face* density criteria for atypical glands were added to previous cross-sectional criteria[1], in accordance with previous *en face* OCT studies[2]. The 5mm square area criteria was chosen as a threshold based on assessment of the image data because glands distributed over a larger area in *en face* images appeared to be unassociated. Adjacent areas that had >5 atypical glands were recorded as a single clustered region. The 1mm depth range was chosen to encompass the boundary of lamina propria and muscularis mucosa. AGCs were assessed using simultaneous 'orthoplane' viewing of the *en face* and cross-sectional image series. *En face* IMPs were defined following recent *en face* OCT studies[2] and using criteria similar to narrowband imaging (NBI)[3], including distortion/absence of mucosal patterns. The two features were assessed separately and demarcated in the *en face* images. The features were counted to determine the absolute occurrence rate (occurrences/patient), per-area occurrence rate (occurrences/approximate contacted area), and occurrence rate of IMP with underlying AGC. Features occurring in BE near (within 1cm of the GEJ) were noted.

Results

Supplementary Figure 1 shows representative results from a treatment-naïve patient referred for treatment for prior histological LGD diagnosis. The area imaged (Figure 2A) was ~4cm (capsule circumference) x ~9cm (approximate pullback length). The mean tissue contact over the en face BE region of analysis was 98%. The capsule moved slower/faster than the tether pullback over some regions, resulting in a stretched/compressed appearance because the en face image is displayed versus time, rather than the actual longitudinal distance traveled by the capsule. Despite these artifacts, features in the en face images are distorted only in the longitudinal direction and can be interpreted by experienced readers. AGCs (Figure 1B-C) were observed in the en face images. A contrast-enhanced, full depth projection en face OCT image (Figure 1D) was used to visualize regions of regular mucosal patterns (Figure 2G), IMPs (Figure 1E-F), and absence of pattern (interpreted as irregular) (Figure 1H). Some IMPs had underlying AGCs. Cross-sectional images also showed atypical glands (Figure 1I-J), as well as features such as surface signal > subsurface (Figure 1K), and normal columnar epithelium (Figure 1L).

Supplementary Table 1 and Figure 2 summarize the features observed in the patient cohorts (history of NDBE vs dysplasia/treatment). The occurrence rates of IMPs, AGCs, and overlapping (combined) features are reported on a per-patient and per-area basis. In the neoplasia cohort, IMPs with underlying AGCs had 1.7 occurrences/patient and 0.36 occurrences/area, while in the NDBE cohort there were only 0.22 occurrences/patient and 0.01 occurrences/area.

The occurrence rate for the NDBE cohort is reported by patient history at the time of imaging, noting that 2/9 patients in the NDBE cohort each had a single occurrence of IMP with underlying AGC; standard endoscopic Seattle protocol of 4-quadrant biopsies obtained from the longitudinal position closest to the observed features showed LGD in one patient and indefinite dysplasia in the second patient. Thus, the rate of IMP with underlying AGC in the 7 patients with true NDBE status was zero occurrences/patient. These results suggest a strong association of IMP and underlying AGC with neoplasia, albeit in a small study cohort.

Discussion

The occurrence rates of IMPs and AGCs in this study were associated with neoplasia status and treatment history. IMPs with underlying AGCs occurred more frequently in treatment naïve neoplasia patients (3/3 patients, 3.0 occurrences/patient, 0.31/area) than in the treated neoplasia cohort (2/4 patients, 0.75 occurrences/patient, 0.40/area). 2/9 patients in the history of NDBE cohort had no endoscopically visible lesions and received only Seattle protocol, but OCT imaging revealed IMPs with underlying AGCs. These patients were subsequently found to have LGD and indefinite dysplasia on 4-quadrant biopsies from the approximate longitudinal position of these OCT features. The remaining 7 patients with NDBE history did not have IMPs with underlying AGCs and did not have dysplasia on Seattle protocol biopsy. These findings suggest that IMPs with underlying AGCs are a candidate marker for dysplasia and elevated risk.

These results are consistent with a previous endoscopic study using micromotor OCT probes with biopsy or endoscopic mucosal resection (EMR) histology from the OCT imaged region[2]. Micromotor probes image a small (~10mm x 16mm) field-of-view compared to the tethered capsule (~4cm x 10cm). However, spatially correlated biopsy/resection can be performed using a dual-channel endoscope. In the previous study, blinded reading of 74 OCT datasets with correlated histology (49 NDBE, 25 neoplasia) from 44 patients identified atypical glands under IMPs in 75% of neoplasia (96% of treatment-naïve neoplasia) vs. 30% of NDBE (43% of short- and 18% of long-segment NDBE)[2].

Atypical glands can be assessed using either *en face* or cross-sectional views, but the extent and organization of clustering are more readily appreciated in *en face* views. Atypical glands are associated with neoplasia, but also occur in NDBE, reducing specificity for this feature[4]. The present study showed AGCs occurred at rates of 3.0/patient and 0.60/area in neoplasia, versus 1.3/patient and 0.30/area in patients with NDBE history. A multi-center study of NBI criteria for dysplasia based on *en face* features including IMPs showed 80% sensitivity and 88% specificity[3]. Either AGCs or IMPs can have substantial independent occurrences, but their combination (AGC under IMP) may have a sensitive and more specific association with neoplasia. The combination of OCT atypical gland criteria with *en face* IMPs from either NBI or OCT may improve the detection of neoplasia vs NDBE.

A substantial fraction of features in all cohorts occurred near the GEJ (Supplementary Table 1), a region reported to have low OCT feature specificity[2 , 5]. Glands in the proximal cardia may be mistaken for atypical glands near the GEJ and result in over-counting / false positives[6]. The role of cardiac mucosa at the GEJ is debated, and it has been proposed that GEJ glands may progress to BE and have pre-malignant potential[7]. The large field-of-view and *en face* visualization provided by tethered capsules may improve GEJ assessment compared to slower cross-sectional OCT balloon imaging and elucidate the pathogenesis of BE and dysplasia originating from GEJ features. Treated patients appeared to have higher rates of IMPs and AGCs. This finding might be associated with the ablation procedure, however, these patients had short residual BE lengths after prior treatment; therefore the majority of features were near the GEJ, where glands in the proximal cardia might be mistaken for AGCs.

Limitations of this analysis include the lack of histological correlation with OCT features. 4-quadrant biopsies were performed in patients with NDBE history, but patients with prior treatment-naïve dysplasia referred for treatment were not systematically biopsied. Two patients in the NDBE history cohort each had a single occurrence of IMP with underlying AGC, with biopsy showing dysplasia and indefinite for dysplasia respectively. Furthermore, our results are consistent with a recent study using micromotor OCT probes which had correlated biopsy/EMR histology[2]. Our tethered-capsule did not have the capability to mark regions for biopsy or directly obtain biopsy. However, laser marking using cross-sectional image guidance has been reported in tethered capsules[8], and future capsule studies could adapt the laser marking paradigm to guide biopsy using both *en face* and cross-sectional features. Alternatively, capsule imaging could be performed during EGD, and NBI could be used to identify IMPs with associated sub-surface OCT features for biopsy correlation. Another limitation was the small patient enrollment, which necessitated the use of an unblinded OCT expert to interpret the data while cognizant of patient history. However, to our knowledge, this study maps the esophageal mucosa over the widest area with the highest resolution to date. The wide imaging field and high resolution provide new information on candidate features of dysplasia. The findings are therefore important for generating hypotheses that can be tested in future prospective studies with larger enrollments.

| OCR features in patient subgroups | All neoplasia (n=7) | Treatment-naïve neoplasia (n=3) | Ablated neoplasia (n=4)[†] | History of NDBE (n=9) | NDBE (n=7) [††] |
|---|---|---|---|---|---|
| **Irregular mucosal patterns (all)** | | | | | |
| No. of patients | 6/7 | 3/3 | 3/4 | 6/9 | 4/7 |
| Occurrences | 17 (2.4/pt) | 11 (3.7/pt) | 6 (1.5/pt) | 10 (1.1/pt) | 8 (1.1/pt) |
| >1cm from GEJ | 3 (0.43/pt) | 2 (0.67/pt) | 1 (0.25/pt) | 7 (0.77/pt) | 5 (0.71/pt) |
| Occurrences per $cm^2$ area* | 0.56 | 0.40 | 0.72 | 0.06 | 0.05 |
| **Irregular mucosal patterns w/ atypical gland clusters** | | | | | |
| No. of patients | 5/7 | 3/3 | 2/4 | 2/9[††] | 0/7 |
| Occurrences | 12 (1.7/pt) | 9 (3.0/pt) | 3 (0.75/pt) | 2 (0.22/pt) | 0 (0/pt) |
| >1cm from GEJ | 3 (0.43/pt) | 2 (0.67/pt) | 1 (0.25/pt) | 2 (0.22/pt) | 0 (0/pt) |
| Occurrences per $cm^2$ area | 0.36 | 0.31 | 0.40 | 0.008 | 0 |
| **Atypical gland clusters (all)** | | | | | |
| No. of patients | 7/7 | 3/3 | 4/4 | 7/9 | 5/7 |
| Occurrences | 21 (3.0/pt) | 11 (3.7/pt) | 10 (2.5/pt) | 12 (1.3/pt) | 10 (1.4/pt) |
| >1cm from GEJ | 6 (0.86/pt) | 4 (1.3/pt) | 2 (0.5/pt) | 6 (0.67/pt) | 4 (0.57/pt) |
| Occurrences per $cm^2$ area | 0.60 | 0.40 | 0.81 | 0.30 | 0.38 |

*Mean of occurrences/area for each patient, where total area per patient is approximated by BE maximal extent × capsule circumference × fraction of BE area in contact.

[†]One patient out of 4 had no visible BE on endoscopy and was thus excluded from the mean occurrences/area computation. Features observed at the GEJ were counted as occurrences.

[††]The two patients with a history of NDBE having irregular mucosal patterns with underlying atypical gland clusters had biopsies with low-grade dysplasia and indefinite for dysplasia. This column reflects true NDBE status at the time of imaging.

Supplementary Table 1. Occurrence rates of OCT features in BE, irregular mucosal patterns, atypical gland clusters, and irregular mucosal patterns with underlying atypical gland clusters in patient subgroups.

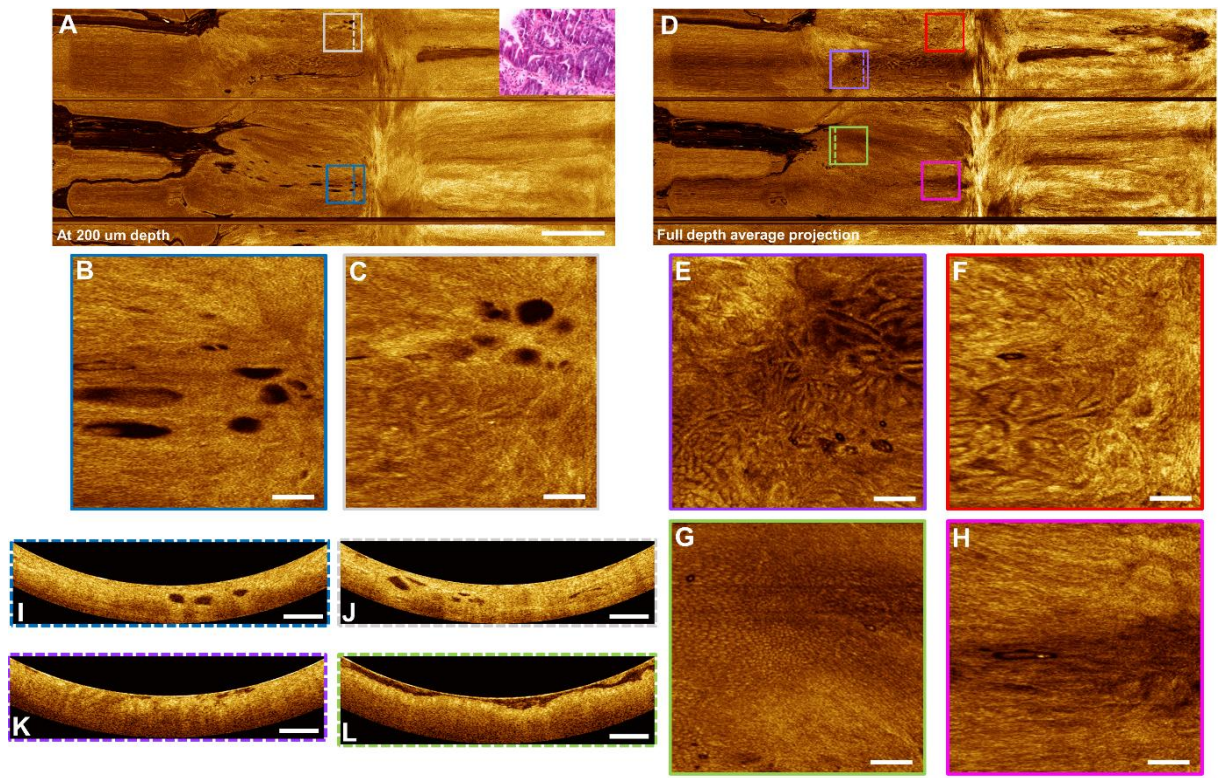

Supplementary Figure 1. Tethered capsule volumetric OCT from a patient with C0.5M2 and treatment-naïve low-grade (basal crypt) dysplasia with biopsy (inset) from previous EGD. *En face* OCT image (A) at 200 um depth from surface, averaged over 80 um depth range (200 um to 280 um), showing dilated glands and (D) *en face* image averaged over ~1 mm depth range (full projection) for contrast enhancement of mucosal pattern. Scale bar 1 cm. (B, C) Atypical gland clusters. (E, F) Irregular mucosal pattern. (G) Regular mucosal pattern. (H) Absence of mucosal pattern, interpreted as irregular. (I-L) Cross-sectional images co-registered to *en face* regions of interest. (I) Shows 3 dilated glands with atypical shape, but does not show the substantial clustering of >5 glands seen in the *en face* image. (K) Shows surface signal higher than subsurface, but does not show the mucosal pattern irregularity seen in the *en face* image. (L) Shows loose contact at the gastroesophageal junction, which may confound differentiation between gastric and BE tissue. All other scale bars 1 mm.

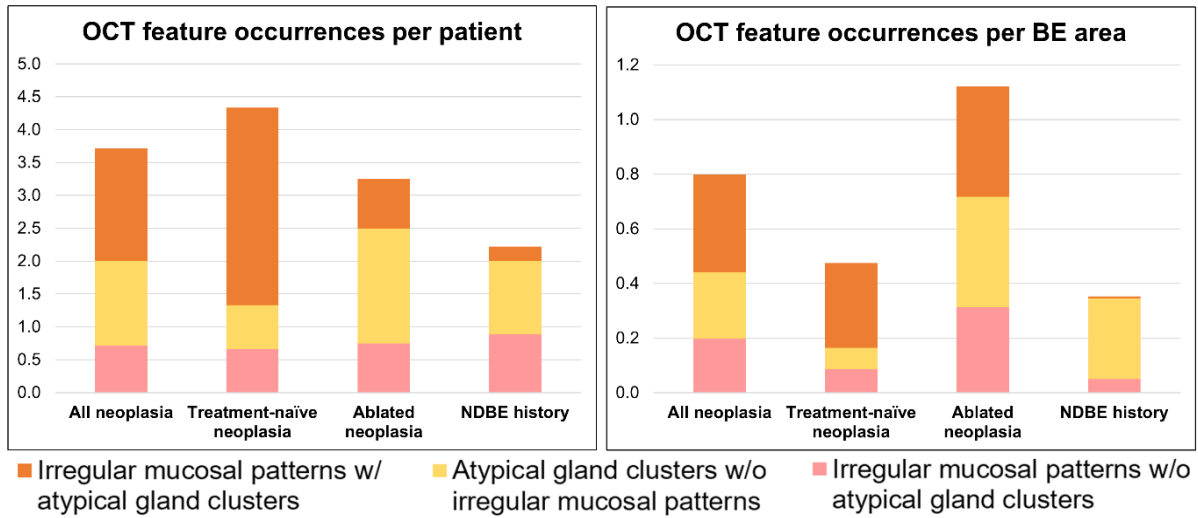

Supplementary Figure 2. Stacked bar charts of OCT feature occurrences per patient and per cm$^2$ of BE area. The features are plotted to avoid repeat counting, such that the stacked height indicates the total feature count. Neoplasia patients had a high occurrence rate of irregular mucosal patterns with underlying atypical gland clusters, while patients with history of NDBE showed a much lower occurrence rate. Atypical gland clusters were numerous in all subgroups and associated with a proximity to the GEJ (Supplementary Table 1).

Supplementary Video 1: Demonstration of orthoplane viewing of *en face* and cross-sectional OCT image series, using dataset presented in Figure 2. Ultrahigh-speed tethered capsules acquire 1,000,000 A-scans per second with 300 images/second at a pullback speed of ~1 cm/second. Ultrahigh speed is important for *en face* OCT because each pixel in the *en face* view requires one A-scan. Orthoplane viewing is similar to CT reading and enables rapid and comprehensive assessment of features.